\documentclass[12pt]{iopart}

\usepackage{graphicx} 
\usepackage{color}
\usepackage{ctable}
\usepackage{multicol}
\usepackage{multirow}
\usepackage{caption}
\newcommand {\be}{\begin{equation}}
\newcommand {\ee}{\end{equation}}
\newcommand {\ba}{\begin{eqnarray}}
\newcommand {\ea}{\end{eqnarray}}

\newcommand{\al}{\lambda_{_L}}
\newcommand{\ac}{\lambda_{_C}}
\newcommand{\ar}{\lambda_{_R}}
\newcommand{\massl}{m_{_L}}
\newcommand{\massc}{m_{_C}}
\newcommand{\massr}{m_{_R}}
\newcommand{\nl}{n_{_L}}
\newcommand{\nc}{n_{_C}}
\newcommand{\nr}{n_{_R}}
\newcommand{\nlr}{n_{_{L,R}}}
\newcommand{\To}{T_{_0}}
\newcommand{\DT}{\Delta T}
\newcommand{\Tmpl}{T_{_L}}
\newcommand{\Tmpr}{T_{_R}}

\begin{document}

\title[]{Thermal rectification in mass-asymmetric one-dimensional anharmonic oscillator lattices with and without a ballistic spacer}

\author{M.~Romero-Bastida and Brandon Armando Mart\'{\i}nez-Torres}
\address{SEPI ESIME-Culhuac\'an, Instituto Polit\'ecnico Nacional, Av. Santa Ana No. 1000, San Francisco Culhuac\'an, Culhuac\'an CTM V, Coyoac\'an, CDMX 04440, Mexico}
\ead{mromerob@ipn.mx}

\date{\today}

\begin{abstract}
In this work we perform a systematic analysis of various structural parameters that have influence on the thermal rectification effect, i.e. asymmetrical heat flow, and the negative differential thermal resistance ---reduction of the heat flux as the applied thermal bias is increased--- present in a one-dimensional, segmented mass-graded system consisting of a coupled nearest-neighbor harmonic oscillator lattice (ballistic spacer) and two diffusive leads (modeled by a substrate potential) attached to the lattice at both boundaries. At variance with previous works, we consider the size of the spacer as smaller than that of the leads. Also considered is the case where the leads are connected along the whole length of the oscillator lattice; that is, in the absence of the ballistic spacer. Upon variation of the system's parameters it was determined that the performance of the device, as quantified by the spectral properties, is largely enhanced in the absence of the ballistic spacer for the small system-size limit herein considered.
\end{abstract}

\pacs{44.10.+i, 05.60.-k, 05.10.Gg, 07.20.-n}


\section{Introduction\label{sec:Intro}}

Heat conduction in low-dimensional systems has become the subject of a large number of theoretical and experimental studies in recent years~\cite{Abhishek08a,Lepri16}. A large part of this interest has been to a large extent driven by the intense study of the thermal rectification (TR) effect, which manifests itself as the asymmetry of heat current when the temperature difference is inverted. It plays a pivotal role in the thermal management at nanoscale and in the development of nanoscopic-sized thermal management devices and functional materials, which certainly explains the large number of works devoted to obtain a deeper understanding of this effect~\cite{Roberts11,Li12,Maldovan13,Sklan15}, with the ultimate aim set in practical applications. The advance in the field has been very impressive since only a few years elapsed from the the first theoretical proposal involving a structure consisting of coupled one-dimensional (1D) anharmonic oscillator lattices~\cite{Terraneo02} to the first successful experimental implementations by means of asymmetric nanotube structures~\cite{Chang06a}, of coupled cobalt oxides with different thermal conductivities~\cite{Kobayashi09}, and polycrystalline samples with asymmetric shape~\cite{Sawaki11}. Furthermore, benefiting from the development of nano- and micro-technology in modeling and nanofabrication, it is now possible to engage in both experimental and theoretical research of nanostructures, such as carbon nanotube bundles~\cite{Aiyiti18} and asymmetric 2D graphene monolayers~\cite{Wang17,Zhao22}, with a potential to develop practical devices for intelligent thermal management, novel thermal transistors, and energy harvesting, among others.

While different rectification mechanisms have been identified and discussed, the common feature characterizing such devices is an underlying structural asymmetry along the direction of the heat flux~\cite{Eckmann06,Casati07,Leitner13,Liu14,Reid19}. To this date the most explored mechanism to obtain a structural asymmetry consists in merging two materials exhibiting different heat transport properties, which was first employed in Ref.~\cite{Terraneo02} and subsequently employed to improve the rectification efficiency of similar models~\cite{Li04a,Hu05}. Another strategy consists in employing graded systems, i.e., inhomogeneous systems whose structure changes gradually in space, which have been both theoretically shown to be optimal materials for thermal diodes~\cite{Pereira10b,Pereira11,Wang12} and extensively studied for various structural modifications~\cite{Yang07,Romero13,Romero17}. In both of the aforementioned proposals it was determined that the match or mismatch of spectral properties of the different parts of the system and afforded by the anharmonicity of the employed lattice is also a necessary condition for the appearance of TR.

Recently, a very interesting proposal consisting of a 1D segmented mass-graded harmonic oscillator lattice, with the boundary regions of the system (termed left and right lead) interacting with a substrate modeled by an onsite potential~\cite{Chen18}, was advanced in order to solve the problem of the rapidly decaying rectification efficiency as the system size increases~\cite{Hu06,Hu06a}. It has been shown that the central segment, without interaction with an onsite potential and termed \emph{ballistic spacer}, contributes crucially to remove dependence of rectification on the system size. This result seem to be quite robust upon variation of the system parameters, as well as to the presence of anharmonic~\cite{Chen18} and next-nearest-neighbor interactions among the oscillators~\cite{Romero21}. 

Now, in the aforementioned rectifier the crucial component, the ballistic spacer, has a larger size than that of the leads. However, the opposite case wherein the leads have larger dimensions than the central spacer has not yet been explored. The relevance of studying this case stems from the fact that, notwithstanding the importance of a sizable rectification in the thermodynamic limit, there are some instances wherewith it could be important to have an efficient nanoscopic-sized rectifier. For example, in standard molecular junctions a molecular structure is placed between and connected to conducting substrates (leads) which are usually metals; this arrangement has been studied both experimentally~\cite{Cui19} and by numerical simulation~\cite{Sharony20,Dinpajooh22}. Thus, when large metal leads and short hydrocarbon molecules are employed the resulting system can be considered as an instance of a possible implementation of the considered oscillator model with a short ballistic spacer compared to the longitudinal dimensions of the leads. Since it has been established, both theoretically and experimentally, that heat transport through short hydrocarbon and similar chain molecules is ballistic~\cite{Ness17,Rubtsov19}, one could consider the hydrocarbon molecule as a ballistic spacer, and apply asymmetry either along the molecule's length or on the metal leads in order to obtain the TR effect.

In this work we will consider the aforementioned modification of the rectifier model considered in~\cite{Chen18}, i.e., a ballistic channel of smaller dimensions than the leads in the boundaries. Furthermore, we will also consider the limit wherein there is no ballistic channel at all, which entails considering a segmented mass-graded harmonic lattice connected to an inhomogeneous substrate. This alternative configuration will allow us to compare the rectification efficiency of both systems, with and without a ballistic spacer, in the small system-size limit. For both instances we will also perform a detailed analysis of the structural factors that can have an influence on the negative differential thermal resistance (NDTR) effect ---namely, the larger the temperature difference, the less the heat flux through the system---, which is crucial for the design of novel nanoscopic devices such as thermal transistors, logic gates, and memories~\cite{Li12,Sklan15}. Among these are the magnitude of the parameters that quantify the strength of the onsite potential, as well as the magnitude of both the largest mass and temperature differences.

This work is organized as follows: In Sec.~\ref{sec:Model} we present the model and the relevant details of its numerical implementation. Our results for the rectification and NDTR effect are presented in Sec.~\ref{sec:S2}. Our conclusions and final observations are summarized in Sec.~\ref{sec:Disc}.

\section{The model and details of numerical simulation\label{sec:Model}}

\begin{figure}\centering
\includegraphics[width=0.69\linewidth,angle=0.0]{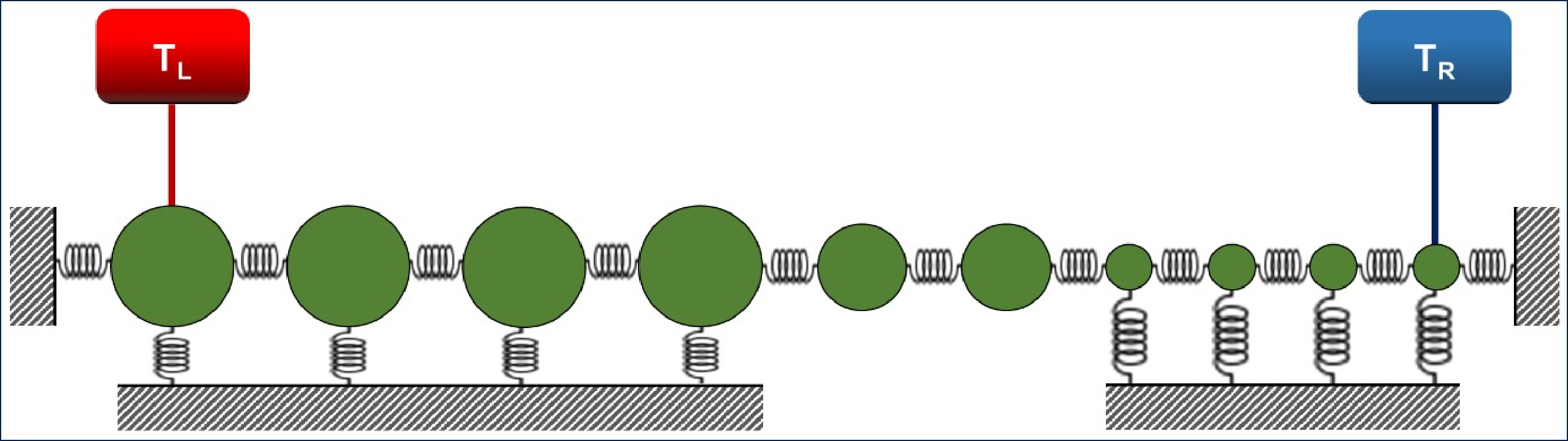}
\caption{Schematic representation of a lattice of coupled nearest-neighbor oscillators interacting with an external substrate and coupled at both ends with two thermal reservoirs working at different temperatures. The central segment, composed of $\nc$ oscillators, is attached at the left (right) end to a lead composed of $\nl$ ($\nr$) oscillators, which in turn interact with a substrate, modeled by an onsite potential of strength $\al$ ($\ar$). $\nlr>\nc$ will always be maintained.}
\label{fig:0}
\end{figure}

A schematic setup of the 1D system to be studied in the following is presented in Fig.~\ref{fig:0}. It is a lattice of $N$ oscillators, with separation $a$ in their equilibrium position, coupled by the nearest-neighbor harmonic potential $V(x)=k_{_0}x^2/2$, where $k_{_0}$ is the harmonic constant. For a harmonic lattice in contact with a heat reservoir specified by a temperature $T$ there are four independent parameters $m$, $a$, $k_{_0}$, and $k_{_B}$, where the latter denotes the Boltzmann constant. Since the dimensions of all the physical quantities involved in heat transport can be expressed by the proper combination of these four parameters, one can introduce dimensionless variables measuring lengths in units of $[a]$, momenta in units of $[a(mk_{_0})^{1/2}]$, temperature in units of $[k_{_0}a^2/k_{_B}]$, frequencies in units of $[(k_{_0}/m)^{1/2}]$, energy and the amplitudes of the onsite potentials to be defined below in units of $[k_{_0}a^2]$, and heat fluxes in units of $[a^2k_{_0}^{3/2}/m^{1/2}]$. In this new set of variables the harmonic potential now reads as $V(x)=x^2/2$. Next, $\nl$ ($\nr$) oscillators with mass $\massl$ ($\massr$) in the left (right) side of the system interact with substrates of negligible thermal conductivity. This interaction is quantified by a quartic, $\phi^4$ onsite potential $U_{_{L,R}}(x)=\lambda_{_{L,R}}x^4/4$, being $\lambda_{_{L,R}}$ the parameters that quantifies the strength of the anharmonic contribution of the onsite potential on each side of the system. Therefore, the two anharmonic leads are connected by a purely harmonic lattice, i.e., a ballistic channel which corresponds to an onsite potential strength of $\ac=0$, composed of $\nc$ oscillators of mass $\massc$; thus, the total system size can be written as $N=\nl + \nc + \nr$. In the following we will take values for these latter variables as $\nlr>\nc$, which corresponds to a ballistic spacer of smaller length than that of the anharmonic leads. Then the equations of motion for each lattice oscillator can be written as $\dot q_i =p_i/m_i$ and
\ba 
\dot p_i&=&F(q_i-q_{i-1}) - F(q_{i+1}-q_i) - \sum_{j=1}^{\nl}\delta_{ij}\al q_j^3 - \!\!\!\!\!\sum_{j=N-\nr+1}^{N}\!\!\!\!\!\!\delta_{ij}\ar q_j^3\cr
   & + & \delta_{1i}\,(\xi_{_1} - \gamma_{_L} p_{_1}) + \delta_{Ni}\,(\xi_{_N} - \gamma_{_R} p_{_N}),
\ea
where $\{m_i,q_i,p_i\}_{i=1}^N$ are the dimensionless mass, displacement, and momentum of the $i$th oscillator. $F(x)=-\partial_x V(x)$ is the harmonic inter-oscillator force. Fixed boundary conditions are assumed ($q_{_0}=q_{_{N+1}}=0$), which are the same as those employed for the molecular junction studied in Ref.~\cite{Sharony20}. Physically the leads constitute the interface with infinitely large thermal reservoirs with negligible motion, a fact that motivates the use of the aforementioned boundary conditions. Furthermore, the smallest system size of $N=28$ atoms therein considered can be taken as an example of our proposal since an organic molecule of $8$ atoms is placed between two leads, composed of $10$ gold atoms each. Next, $\xi_{_{1,N}}$ are independent Wiener processes with zero mean and variance $2\gamma_{_{L,R}}T_{_{L,R}}m_{_{1,N}}$, being $\gamma_{_{L,R}}$ the coupling strength between the first (last) oscillator in the lattice and the left (right) reservoir operating at temperature $T_{_L}$ ($T_{_R}$). Since the main objective of the present work is to study the influence of structural asymmetries on TR we will hereafter consider exclusively symmetric coupling strengths of fixed magnitude, i.e., $\gamma_{_{L,R}}=0.5$ in all considered instances; the effect of asymmetric coupling strengths was thoroughly considered in Ref.~\cite{Romero20}. We can define the average temperature $\To\equiv(\Tmpl+\Tmpr)/2$ and difference $\DT\equiv\Tmpl-\Tmpr$; thus, $T_{_{L,R}}=\To\pm\DT/2$. Hereafter we will consider a mass distribution given by the mass values $\massl>\massc=\massl/2>\massr=1$, which amounts to a discontinuous left-to-right mass gradient. The aforementioned equations were integrated with a symmetrical stochastic Verlet integrator implemented in an in-house Fortran code with a time step of $10^{-2}$ in all considered cases.

Once the non-equilibrium stationary state is attained, the total heat flux $J$ is computed as the algebraic average of $J_i=\langle\dot q_i F(q_{i+1}-q_i)\rangle$, where $\langle\cdots\rangle$ indicates time average, over the $N-2$ unthermostated (bulk) oscillators. This compact expression is derived from the symmetrical one directly obtained from the discretization of the continuity equation for this system by exploiting the equality $\langle\dot V(q_{i+1}-q_i)\rangle=0$ that holds in the stationary state~\cite{Lepri03}. The heat flux can be computed, employing the aforementioned time step, with a precision of $\mathcal{O}(10^{-6}-10^{-8})$. By $J_+$ we denote the heat flux when the high temperature reservoir is attached to the heavy loaded end of the system and by $J_-$ the flux when that same reservoir is now connected to the opposite end of the lattice, i.e., the positions of the reservoirs are interchanged. With the quotient $r\equiv|J_+/J_-|$ we quantify the rectification efficiency of this device.

In the original reference that introduced the herein employed model~\cite{Chen18} there is no information whatsoever as to the relative contributions to TR of the nonuniform mass distribution and the asymmetry of the amplitudes of the onsite potentials. Therefore we performed some simulations with $\To=0.1$, $\Delta T=0.16$, and $\nlr=16$ for two $\nc$ values: 0 and 8. For $\nc=0$ one simulation had parameters to assess the influence of a segmented mass distribution and uniform amplitude of the onsite potential; a second one had parameters to isolate the effects of inhomogeneous onsite potential amplitude with uniform mass distribution; finally, a third one had those which correspond to the values employed in this work and in previous ones~\cite{Chen18,Romero20}. For $\nc=8$ the same procedure was performed. The complete list of parameter values for each $\nc$ case, together with the resulting rectification values, are reported in Table~\ref{tab:table1}. These results clearly indicate that the nonuniform mass distribution is the main origin of the rectification effect and that, when additionally the amplitudes of the onsite potential on each side are given nonuniform values, then a major boost on the rectification value is obtained, more for the $\nc=0$ instance. Therefore the nonuniformities on both mass distribution and onsite amplitudes have to be taken into account simultaneously to obtain the desired TR efficiency. The direction of the mass gradient is justified by recalling that heat flow diminishes in the direction of increasing mass density~\cite{Yang07}, which is precisely the situation in the reverse-bias configuration. Furthermore, a larger amplitude of the onsite potential in the right side, together with the condition $T_{_L}<T_{_R}$, results in a further decrease in the heat flow and an increase in the ensuing rectification value.

\begin{table}[h]\centering
\caption{\label{tab:table1} Values of the structural parameters for simulations with $\nc=0$ and $\nc=8$.}
\begin{tabular}{|ccccc|cccccc|}
\hline
\multicolumn{5}{|c|}{$\nc=0$} & \multicolumn{6}{c|}{$\nc=8$} \\
\hline
$\massl$ & $\massr$ & $\al$ & $\ar$ & $r$ & $\massl$ & $\massc$ & $\massr$ & $\al$ & $\ar$ & $r$ \\
\hline
10  & 1 & 1 & 1 & 4.71 & 10 & 5 & 1 & 1 & 1 & 3.75 \\
1   & 1 & 1 & 5 & 1.2  & 1  & 1 & 1 & 1 & 5 & 1.1  \\
10  & 1 & 1 & 5 & 16.5 & 10 & 5 & 1 & 1 & 5 & 10.6 \\
\hline
\end{tabular}
\end{table}

\section{Thermal rectification and NDTR\label{sec:S2}}

\begin{figure}\centering
\includegraphics[width=0.49\linewidth,angle=0.0]{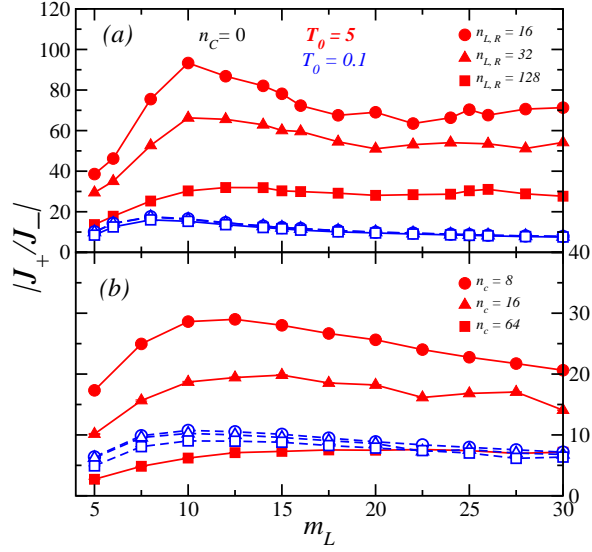}
\caption{(a) Thermal rectification $r$ vs $\massl$ with $\nc=0$. Open symbols correspond to $\To=0.1$ and $\Delta T=0.16$, whereas filled ones to $\To=5$ and $\Delta T=9$. Circles correspond to $\nlr=16$ ($N=32$), triangles to $\nlr=32$ ($N=64$), and squares to $\nlr=128$ ($N=256$). (b) Same as (a) but now for $\nc=8$ (circles, $N=40$), $\nc=16$ (triangles, $N=80$), and $\nc=64$ (squares, $N=320$). $\al=1$ and $\ar=5$ in all instances. Error bars are smaller than symbol size. Lines are a guide to the eye.}
\label{fig:1}
\end{figure}

In Fig.~\ref{fig:1}(a) we plot $r$ vs $\massl$ values, all with $\al=1$ and $\ar=5$, to assess the effect of the mass asymmetry in the TR efficiency of these lattices in the absence of a ballistic spacer, i.e. $\nc=0$, with $N=\nl+\nr$. The simulation times were of $2\times10^7$ and $\sim6\times10^7$ time units for the transient and stationary time intervals, respectively. We first notice that, for the case without a ballistic spacer, the greatest rectification efficiency is obtained, in the high-temperature case, for the smallest system size of $N=32$, i.e. $r=93$, at a mass value of $\massl=10$. The rectification steadily decreases as the system size increases, until at $N=256$ the rectification value $r=32$ becomes almost mass-independent for $\massl>10$ values. For the low-temperature instance the rectification figures are consistently lower than those at high temperature, as expected, but are largely independent of the system size for the $N$ values considered. The maximum TR efficiency is now obtained with $\massl=8$. For the case with a ballistic spacer depicted in Fig.~\ref{fig:1}(b) there are some differences worth remarking, being the most immediate that the rectification values are lower than those obtained without a ballistic spacer. In the high-temperature instance the highest rectification figure, obtained for $\massl=12.5$ and $\nc=8$, has a much slower and smoother decrease for higher $\massl$ values than the corresponding cases reported in panel (a). This feature persists for higher system size values, except that the maximum rectification figure now corresponds to $\massl=15$ and $22.5$ for the $\nc$ values of $16$ and $64$, respectively. But more important, it is remarkable that there is a sharp reduction of $r$ as $N$ increases. This result stands in sharp contrast to those wherein the ballistic spacer is larger than the end leads, where it was observed that there is no such size dependence~\cite{Chen18,Romero21}. The low-temperature instance also presents some interesting features as well. First, the aforementioned reduction of the rectification value for larger system sizes is also herein observed, but much reduced in magnitude. The maximum rectification figure is obtained at $\massl=10$ for all system sizes considered. Finally, for $\massl<20$ values the rectification obtained for all system sizes considered in the low-temperature regime is actually larger than that of the $N=256$ instance for high temperature. Thus there is a range of mass values that are entirely feasible to obtain experimentally wherein moderate rectification figures, i.e. $r\sim10$, can be obtained almost independently of the system size. Now although the larger $\massl$ values herein considered can be difficult to obtain experimentally, it is viable that another, hitherto unexplored but experimentally feasible, asymmetry could be implemented in this or related models, as in the fluid system with asymmetric contact areas with the reservoirs to obtain TR proposed in Ref.~\cite{Komatsu10}.

\begin{figure}\centering
\includegraphics[width=0.59\linewidth,angle=0.0]{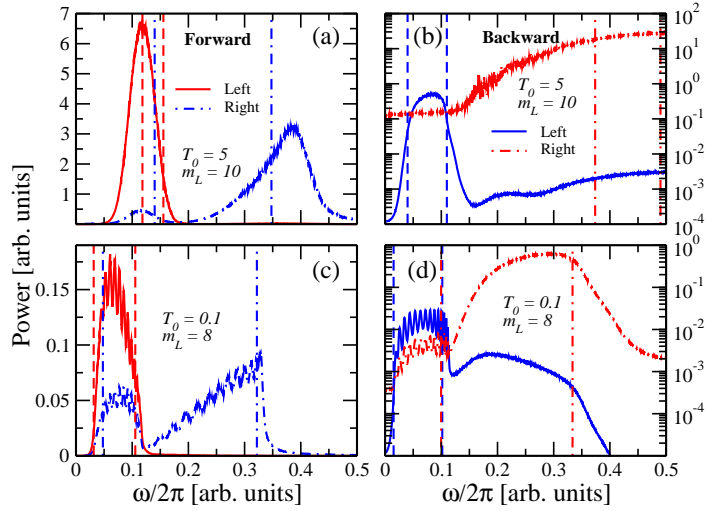}
\caption{Power spectra of two oscillators ($i=16$ and $17$) on each side of a lattice for two temperature values, $\To=5$ (a,b) with $\massl=10$ and $\To=0.1$ (c,d) with $\massl=8$; the corresponding $\Delta T$ values are $9$ and $0.16$ respectively. $\nlr=16$, $\nc=0$, $\al=1$, and $\ar=5$ in all instances. Forward bias configuration corresponds to panels (a,c) and reverse bias to panels (b,d). In each panel vertical solid and dashed lines indicate the lower and upper limits of the left and right segment phonon bands, respectively. See text for details.}
\label{fig:2}
\end{figure}

In order to gain further insight into the origin of the aforementioned behavior we compute the power spectra (PS) $P_i(\omega)=\langle|\tau^{-1}\!\!\int_{_0}^{\tau}\!\! dt\dot q_i(t)\exp(-\mathrm{i}\omega t)|^2\rangle$ of two oscillators ($i=16$ and $17$) at each side of the boundary for the case without a ballistic spacer $\nc=0$ and $\nlr=16$; the Fourier transform is computed over an interval of $\tau=2^{12}$ time units and $\langle\cdots\rangle$ indicates an average over the complete stationary time interval. The results are reported in Fig.~\ref{fig:2}. For the $\massl=10$ case ---which corresponds to the highest rectification according to Fig.~\ref{fig:1}(a)--- at high temperature depicted in Fig.~\ref{fig:2}(a), the spectrum corresponding to the left side of the system in the $J_+$ configuration lies within the low-frequency region whereas for the one corresponding to the right side the spectral power is more concentrated on the high-frequency region, being this phenomenology consistent with the asymmetries of the onsite potential amplitudes and mass distribution on each side. It has been shown that the heat carriers determining the heat transport in nonlinear lattices are the renormalized phonons~\cite{Nianbei10}, i.e., phonons with a dispersion relation $\hat\omega_k=(\omega_k^2+\gamma)^{\frac{1}{2}}$ renormalized with a coefficient $\gamma$ that encodes the information of the nonlinear interaction that depends only on the temperature or the strength of the nonlinearity; more precisely, $\gamma=\sum_i\langle q_i^4\rangle/\sum_i\langle q_i^2\rangle$~\cite{Li13}. From the classical field approach the coefficient can be numerically calculated to computer precision~\cite{Boyanovsky04} and therefore it can be straightforwardly shown that the active phonon frequencies are located within the phonon band $[(1.23 T^{\frac{2}{3}}_{_{L,R}}/m_{_{L,R}})^{\frac{1}{2}},\{(4k_{_0}+1.23\Tmpl^{\frac{2}{3}})^{\frac{1}{2}}\}/m_{_{L,R}}]$, with $k_{_0}=1$ henceforth as explained in Sec.~\ref{sec:Model}. It can be immediately corroborated that the lower and upper limits of the phonon frequencies are in good agreement with the predicted phonon bands also depicted in that panel. The slight mismatch observed is due to the fact that the aforementioned approximation was performed for a homogeneous lattice at a fixed temperature, whereas in our case both halves of the system have different amplitudes of the onsite potential and temperatures. Now, since both spectra are of similar magnitude and there is an overlap in the low frequency region for the forward-bias configuration, heat flow through the system is favored. On the other hand, for the $J_-$ configuration depicted in panel (b) the right spectra presents a significant increase in the contribution from all frequencies, specially a massive one in the high-frequency region, and the predicted phonon bands have no overlap altogether, thus hindering the heat flow and rendering a high $r$ value. For the low-temperature regime the aforementioned behavior is somewhat altered: both spectra have comparable magnitude and present a discrete structure ---specially the left-side one corresponding to $\al=1$, which diminishes their spectral contribution--- and the right one also has intermediate frequencies of sizable magnitude, reducing the heat flux and thus the TR value compared to the corresponding one for the high-temperature regime as was already noticed in Fig.~\ref{fig:1}. In the reverse-bias configuration depicted in panel (d) the predicted phonon bands have negligible overlap, which is reflected in the fact that the frequencies that carry the largest power values of each spectra are located precisely within those phonon bands. Furthermore, the right spectrum presents an spectral contribution overwhelmingly large in the intermediate-frequency range, similar to that in the high-temperature instance depicted in panel (b). All this factors certainly contribute to the reduced rectification compared to the corresponding high-temperature instance.

\begin{figure}\centering
\includegraphics[width=0.59\linewidth,angle=0.0]{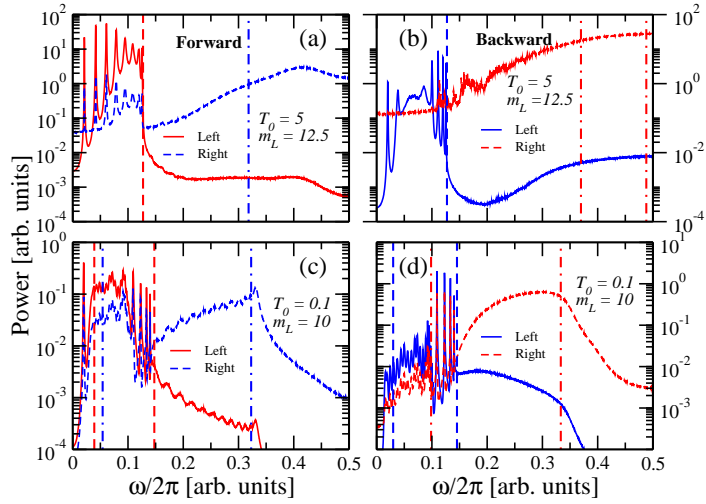}
\caption{Same as in previous figure, but for a lattice with a ballistic spacer. All presented cases correspond to $\nlr=16$ and $\nc=8$. The masses of the left lead are $\massl=12.5$ and $10$ for the high- and low-temperature instances, respectively. The two considered oscillators are those in the right boundary between the ballistic spacer and the right lead ($i=24$ and $25$).}
\label{fig:3}
\end{figure}

The spectral analysis corresponding to the lattice with a ballistic spacer is presented in Fig.~\ref{fig:3} for the case $\nlr=16$ and $\nc=8$ ($N=40$). In the case of the forward-bias configuration for $\massl=12.5$, $\To=5$, and $\Delta T=9$ depicted in panel (a) the part of the bulk (left) spectrum that lies within the low-frequency region has a distinctly discrete structure characteristic of the underlying harmonic dynamics within the ballistic spacer, with a phonon band now given by $[0,(4k_{_0}/\massc)^{{1\over2}}]$, with $k_{_0}=1$. The right (lead) spectrum also has a discrete structure for near-zero frequencies, which indicates that the lead is largely in the harmonic regime and thus the corresponding phonon band is obtained from the aforementioned expression already employed when considering the ballistic spacer. The effect of the onsite potential is revealed in the significant spectral power allocated in the high-frequency region of the right spectrum. The coexistence of harmonic and anharmonic features in the same spectrum is due to the fact that the ballistic spacer and the lead are part of the same system. Within the ballistic spacer the dynamics is largely harmonic (no onsite potential whatsoever) and thus the discrete structure is dominant in the low-frequency region. The anharmonic tail in the high-frequency region is due to its interaction with the lead. In the latter the onsite potential is dominant, and thus the spectral contribution of higher frequencies is stronger, with only a weak discontinuous structure in the low-frequency region compared to that of the spacer. In the reverse-bias configuration presented in panel (b) the phenomenology is almost the same as in the corresponding panel of the previous figure, except that now some low frequencies have an increased spectral power; thus there is an increase of the heat flux in comparison to the case reported in Fig~\ref{fig:2}(b) that results in a lower $r$ figure. For the low-temperature instance, presented in panels (c) and (d) for the $J_+$ and $J_-$ configurations respectively, it can be readily noticed that the anharmonic effects become relevant and thus a low-frequency band-gap is opened, just as in the cases depicted in Fig.~\ref{fig:2}. In the reverse-bias configuration, contrary to the high-temperature instance, there is a strong overlap of the phonon bands, which coincides with the regions of the phonon spectra where the discrete harmonic structure is more noticeable. Therefore the heat flux is increased in the $J_-$ configuration, thus again diminishing the ensuing rectification.

\begin{figure}\centering
\includegraphics[width=0.59\linewidth,angle=0.0]{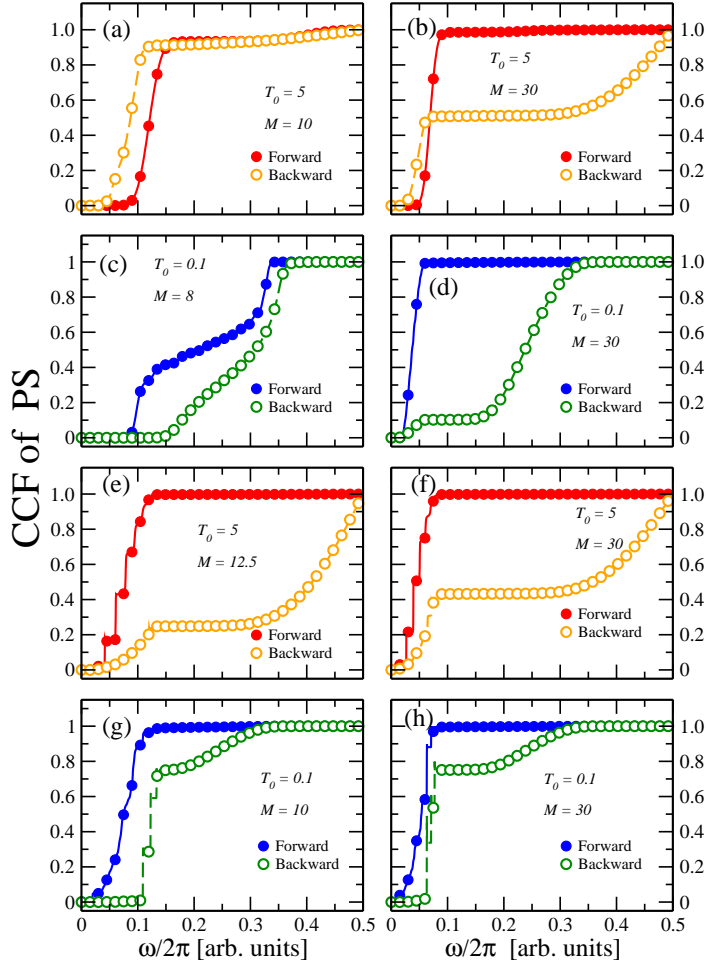}
\caption{(a,b,c,d) CCF of PS between oscillators $i=16$ and $17$ for $\nc=0$. (e,f,g,h) Same as previous panels but for oscillators $i=24$ and $25$ in the right boundary between the ballistic spacer and the right lead for $\nc=8$.}
\label{fig:4}
\end{figure}

To quantify the degree of overlap of the power spectra between oscillators, and thus gain further insight into the mechanisms responsible of TR, the cumulative correlation factor (CCF), introduced in Refs.~\cite{Li05,Zhang17}, is used to represent the match-mismatch degree of vibrational modes among them. The CCF below a specific frequency $\omega_s$ between oscillators $i$ and $j$ is defined as
\be
M_{ij}(\omega_s)={\int_0^{\omega_s}P_i(\omega)P_j(\omega)d\omega\over\int_0^{\infty}P_i(\omega)d\omega\int_0^{\infty}P_j(\omega)d\omega}.
\ee
Each CCF in the two opposite directions is normalized by dividing $M(\omega_s)$ by $M(\infty)$. Previously it has been established that, the more similar the CCFs in the forward and backward directions are, the smaller degree of mismatch of vibrational modes between them is, thus leading to a smaller value of TR~\cite{Dong19}. In the absence of ballistic channel, i.e. $\nc=0$, the result for $\massl=10$ at high temperature, presented in Fig.~\ref{fig:4}(a), indicates that there is a vibrational mismatch in the low frequency-region, which favors the heat flux in the forward direction. For the $\massl=30$ case depicted in panel (b) the mismatch is increased in intermediate- and high-frequency regions. Thus the increased inertia afforded by the larger mass value is associated with a reduction in the vibrational mismatch in the crucial low-frequency region, and thus with a reduced TR efficiency compared to the $\massl=10$ case. For the low-temperature case it can be seen in panels (c) and (d) that the vibrational mismatch is greater compared to the high temperature instances. However, it occurs at intermediate frequencies that are not favorable to heat conduction in either direction, and thus a reduced $r$ value compared to that in the high-temperature case is obtained. This same phenomenology, but sharply increased, is also observed for the $\massl=30$ case, reducing $r$ compared to the $\massl=8$ one. In the $\nc=8$ case for the high $\To$ value, panels (e) and (f), it can be appreciated that there is a higher vibrational mismatch degree at intermediate and higher frequencies compared to low-frequencies, and that there is a lower vibrational mismatch for the high mass value of $\massl=30$ compared to the lower-mass instance. All these factors reduce the rectification as was previously noticed in Fig.~\ref{fig:1}. For the low $\To$ value, panels (g) and (h), there is a small mismatch degree for $\massl=10$ ---much lower than that noticed in the high-temperature instance, panel (e)--- and an even smaller one at $\massl=30$. Thus all these factors explain the low rectification obtained in the presence of a ballistic channel for large mass values.

Next we will explore the effects on the TR efficiency of the amplitude of the onsite potential by varying $\ar$ for $\al=1$ and $\massl=10$; the results are reported in Fig.~\ref{fig:5}. For both cases, with and without a ballistic spacer, it is clear that, in general, an increase in $\ar$ entails an increase in TR, but there are also some important differences worth noting. In the absence of a ballistic spacer it is evident from the results reported in panel (a) that, for the high-temperature regime, $r$ increases as $\ar$ does so, although at a smaller rate as the system size increases, whereas the TR presents a very weak dependence on $\ar$ for all system sizes considered in the low-temperature regime. We also notice that, for all $\ar=1$ instances, the only asymmetry in the lattice is afforded by the segmented mass distribution on it, which certainly explains why the rectification for the high temperature and smaller system sizes are clustered in the range $[40,50]$ in the leftmost part of the figure. In panel (b) we display the results in the presence of a ballistic spacer; it is clear that, in all presented instances, the obtained rectification is much lower than that in the absence of the spacer. Furthermore, the $\nlr=128$ case in the high-temperature regime has lower TR than all the instances at lower average temperature for all considered system sizes. Thus the evidence presented seems to indicate that, for small system sizes, the ballistic spacer has a negative effect on the TR efficiency of the lattice.

\begin{figure}\centering
\includegraphics[width=0.49\linewidth,angle=0.0]{Fig5.eps}
\caption{(a) Thermal rectification $r$ vs $\ar$ with $\nc=0$. Open symbols correspond to $\To=0.1$ and $\Delta T=0.16$, whereas filled ones to $\To=5$ and $\Delta T=9$. Circles correspond to $\nlr=16$, triangles to $\nlr=32$, and squares to $\nlr=128$. (b) Same as (a) but now for $\nc=8$ (circles), $\nc=16$ (triangles), and $\nc=64$ (squares). $\massl=10$ and $\al=1$ in all instances. Error bars are smaller than symbol size. Lines are a guide to the eye.}
\label{fig:5}
\end{figure}

\begin{figure}\centering
\includegraphics[width=0.59\linewidth,angle=0.0]{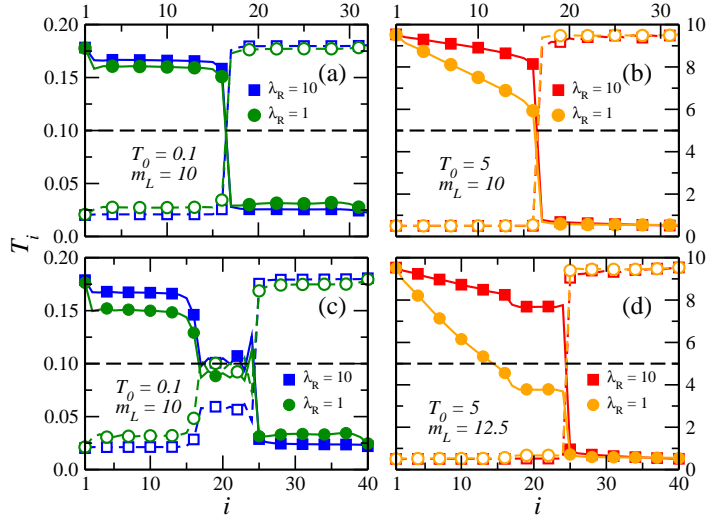}
\caption{Temperature profiles for $\ar=1$ (circles) and $10$ (squares) in the case without ballistic spacer $\nc=0$ for (a) $\To=0.1$ with $\massl=8$ and (b) $\To=5$ with $\massl=10$. Panels (c) and (d) are the same as (a) and (b), but now with $\nc=8$; $\massl=10$ and $\massl=12.5$ for (c) and (d) respectively. Open and void symbols correspond to the forward- and reverse-bias configurations, respectively. $\nlr=16$ in all instances. Error bars are smaller than symbol size.}
\label{fig:6}
\end{figure}

The corresponding temperature profiles of both forward- and reverse-bias configurations for the $\ar=1,10$ instances with and without a ballistic spacer ---$\nc=8$ and $\nc=0$, with $\nlr=16$ in both cases--- for low and high average temperature values are presented in Fig.~\ref{fig:6}. We first notice that, in the low-temperature regime, in both depicted instances the temperature profiles present a high degree of reflection with respect to their corresponding $\To$ values; this behavior is incompatible with a significant rectification figure, as was indeed noticed in the previous figure. Now, for the high-temperature regime, it is clear that both the spatial (along the system length) and reflection (around $\To$) symmetries have been broken, leading to the high rectification values depicted in Fig.~\ref{fig:5}(a). It can also be observed that, for the case with a ballistic spacer shown in panel (d) there is a high degree of variation of the slope of the temperature profile in the forward-bias configuration compared to the $\nc=0$ case of panel (b). This phenomenology leads to a decrease of $J_+(\lambda=1)-J_-(\lambda=10)=12.7\times10^{-3}$ in the former compared to $J_+(\lambda=1)-J_-(\lambda=10)=7.18\times10^{-3} $ in the latter. Such high decrease in $J_+$ value brings the magnitude of both heat fluxes in the forward and reverse-bias configurations closer ($J_-\sim\mathcal{O}(10^{-5}$ in both instances), thus reducing the obtained TR for the case with ballistic spacer compared to that without one.

More information about the previously mentioned asymmetries in the heat flux values can be obtained by studying the interface thermal resistance (ITR) $R_{\pm}\equiv\Delta T/J_{\pm}$, where $\Delta T$ is the temperature difference between the two sides of the interface and $J_{\pm}$ is the heat flux in the forward- and reverse-bias configurations. For the particular cases depicted in Fig.~\ref{fig:6} we have, for $\nc=0$, $\Delta T=T_{16}-T_{17}$ and, for $\nc=8$, $\Delta T=T_{16}-T_{25}$. The ratio $R_{-}/R_{+}$ quantifies the relative magnitude of the ITR in the reverse-bias configuration compared to that in the forward-bias one. It is clear that a larger value of this ratio corresponds to a large TR efficiency. The values of $R_{-}/R_{+}$ for all cases depicted in Fig.~\ref{fig:6} are reported in Table~\ref{tab:table2}. It is clear that $R_{-}/R_{+}$ increases as $\ar$ does so for each reported case. But the highest increment is obtained for the high-temperature instance in the absence of the ballistic spacer. Since the lower values are obtained for the case with a ballistic spacer, it is reasonable to infer that the existence of two interfaces in this case contributes to a decrease in $R_{-}/R_{+}$ and thus to a lower TR value.

\begin{table}[h]\centering
\caption{\label{tab:table2} $R_{-}/R_{+}$ values for each case reported in Fig.~\ref{fig:6}.}
\begin{tabular}{|cccc|cccc|}
\hline
\multicolumn{4}{|c|}{$\nc=0$} & \multicolumn{4}{c|}{$\nc=8$} \\
\hline
\multicolumn{2}{|c|}{$\To=0.1$} & \multicolumn{2}{c|}{$\To=5$} & \multicolumn{2}{c|}{$\To=0.1$} & \multicolumn{2}{c|}{$\To=5$} \\
\hline
\multicolumn{1}{|c|}{$\ar=1$} & \multicolumn{1}{c|}{$\ar=10$} & \multicolumn{1}{c|}{$\ar=1$} & \multicolumn{1}{c|}{$\ar=10$} & \multicolumn{1}{c|}{$\ar=1$} & \multicolumn{1}{c|}{$\ar=10$} & \multicolumn{1}{c|}{$\ar=1$} & \multicolumn{1}{c|}{$\ar=10$}\\
\hline
\multicolumn{1}{|c|}{$3.34$} & \multicolumn{1}{c|}{$19$} & \multicolumn{1}{c|}{$62.41$} & \multicolumn{1}{c|}{$155.14$} & \multicolumn{1}{c|}{$4.6$} & \multicolumn{1}{c|}{$14$} & \multicolumn{1}{c|}{$32.5$} & \multicolumn{1}{c|}{$44.1$}\\
\hline
\end{tabular}
\end{table}

\begin{figure}\centering
\includegraphics[width=0.49\linewidth,angle=0.0]{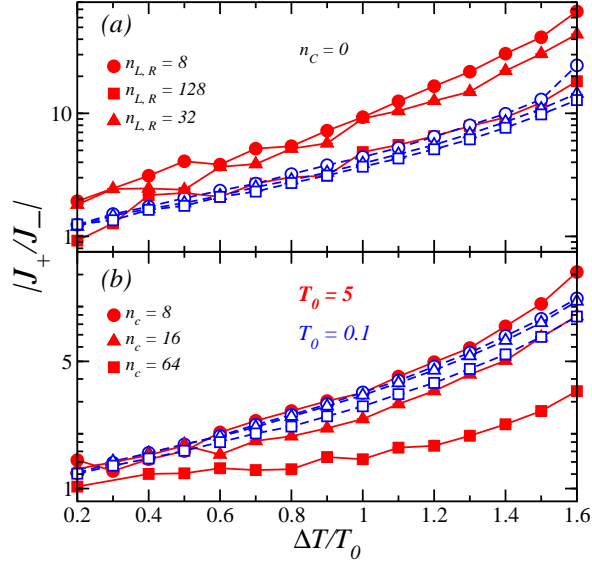}
\caption{(a) Thermal rectification $r$ vs $\Delta T/\To$ for a lattice with $\nc=0$. Open symbols correspond to a $\To=0.1$, whereas filled ones to $\To=5$. Circles correspond to $\nlr=16$, triangles to $\nlr=32$, and squares to $\nlr=128$. For all high-temperature cases, $\massl=10$. In the low-temperature instances with $\nlr=16$, $\massl=8$; for all other $\nlr$ values, $\massl=10$. (b) Same as (a) but now for $\nc=8$ (circles), $\nc=16$ (triangles), and $\nc=64$ (squares). In the high-temperature case for $\nc=16$, $\massl=12.5$; for $\nc=32$, $\massl=15$; for $\nc=64$, $\massl=22.5$. For the low-temperature regime, $\massl=10$ for all $\nc$ values. $\al=1$ and $\ar=10$ in all instances. Error bars smaller than symbol size. Continuous lines are a guide to the eye.}
\label{fig:7}
\end{figure}

Further information on the TR can be obtained if we compare the dependence of the rectification coefficient on temperature difference $\Delta T/\To$ in the presence and absence of the ballistic spacer; the results are reported in Fig.~\ref{fig:7}. In both instances it is clear an exponential-like dependence of $r$ on the imposed temperature difference. For the case of the lattice without a ballistic spacer, panel (a), the high-temperature rectification consistently diminishes as the system size increases for each $\Delta T/\To$ value considered, whereas in the low-temperature regime the aforementioned reduction is largely absent. When the ballistic spacer is present, panel (b), the same phenomenology is observed, but for lower TR values. A difference with the previous case worth noticing is that the results in the high-temperature regime present a much stronger reduction rate when the system size is increased. Next, for $\To=0.1$, as the value of the temperature difference $\Delta T/\To$ increases it is clear that the presence of the ballistic spacer reduces the dependence of $r$ on the system size, just as in the case where the length of the spacer is greater than those of the leads~\cite{Chen18,Romero21}, although the rectification figures are lower than those obtained for the case in which there is no spacer at all.

\begin{figure}\centering
\includegraphics[width=0.49\linewidth,angle=0.0]{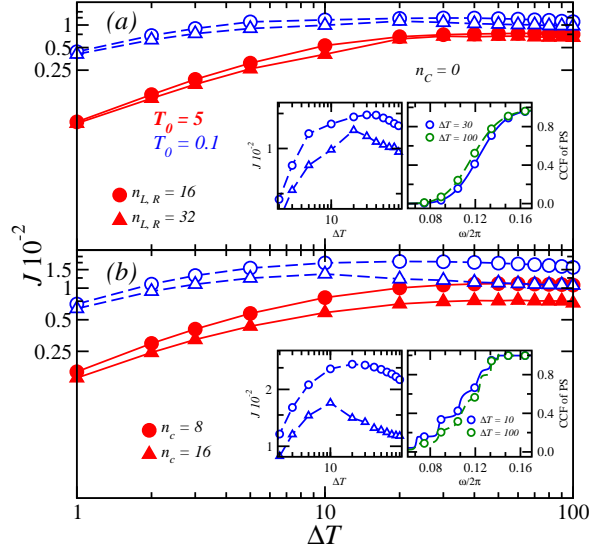}
\caption{(Color online) (a) Heat flux $J$ versus $\Delta T$ for a lattice with $\nc=0$. Open symbols correspond to a $T_{_R}=0.02$ value and filled ones to $T_{_R}=0.5$. Circles correspond to $\nlr=16$ and triangles to $\nlr=32$. (b) Same as (a) but now for $\nc=8$ (circles) and $\nc=16$ (triangles). $\massl=10$, $\massc=5$, $\al=1$, and $\ar=5$ in all instances. The first (left) inset in each panel gives an enlarged view of the NDTR behavior for the $T_{_R}=0.02$ cases and the second (right) one presents the CCFs for the $\Delta T$ value for which $J$ is maximum and that for $\Delta T=100$. Error bars are smaller than symbol size. Continuous lines are a guide to the eye.}
\label{fig:8}
\end{figure}

Next we address an important feature of the segmented mass-graded lattice, namely the NDTR corresponding to a decrease in the heat flux through the system as the temperature difference increases~\cite{Li06}. In Fig.~\ref{fig:8}(a) we plot the heat flux $J$ versus the temperature difference $\Delta T$ for the case without a ballistic spacer for $T_{_R}=0.02$ (open symbols) and $T_{_R}=0.5$ (filled symbols), with $\massl=10$, $\al=1$, and $\ar=5$. When $\Delta T$ is sufficiently small, $J$ and $\Delta T$ are proportional to each other and the system is within its linear response regime. But ---in the case with $T_{_R}=0.02$--- for larger values of $\Delta T$, i.e., in the interval $30<\Delta T<100$ for $\nlr=16$ and in $10<\Delta T<100$ for $\nlr=32$, the system enters a nonlinear response regime where NDTR occurs; in the first inset an enlarged view of the region wherein the effect can be better appreciated is presented. For the case with $T_{_R}=0.5$ the decrease of the heat flux as $\Delta T$ increases, within the interval $70<\Delta T<100$ for $\nlr=16$ and in $60<\Delta T<100$ for $\nlr=32$, is greatly diminished. So far our results are compatible with those reported in Ref.~\cite{He10} for the case of a harmonic lattice with a homogeneous quartic onsite potential of amplitude $\lambda=0.9$ since the decrease $\Delta J\sim9\times10^{-3}$ obtained in the cited work is of similar magnitude to that corresponding to our case for $T_{_R}=0.02$ and $\nlr=32$ of $\Delta J=2\times10^{-3}$. In the second inset it can be noticed that the CCFs for the $\Delta T=30$ and $100$ instances with $\To=0.1$ and $\nlr=16$ are quite similar, with a small mismatch of vibrational modes between them. Next, for the instance with a ballistic spacer displayed in Fig.~\ref{fig:8}(b) NDTR is present, in the case of $T_{_R}=0.02$, for $\Delta T>20$ with $\nlr=16$ and for $\Delta T>10$ with $\nlr=32$; in both instances the reduction of $J$ as $\Delta T$ increases is highly monotonic. On the contrary, for $T_{_R}=0.5$ only a very weak NDTR effect is present for $\Delta T>50$ values for both considered system sizes. Thus the maximum NDTR effect is obtained for $T_{_R}=0.02$ and $\nlr=32$ in the presence of the ballistic spacer since the decrease of $J$ is of $\Delta J=3\times10^{-3}$, compared to $\Delta J=1.5\times10^{-3}$ for the corresponding instance without a ballistic spacer. For this particular instance the CCFs for $\Delta T = 10$ and $100$, reported in the second inset of the panel, have a slightly higher degree of mismatch between them compared to the instance presented in the previous panel, which can account for the stronger NDTR effect in this case. For the last considered case, when an amplitude of $\ar=10$ was taken the decrease in $J$ was of $\Delta T=4\times10^{-4}$ and $\Delta T=10^{-4}$, respectively. A considerable reduction of the NDTR effect was also observed in all other studied instances (not shown) ---and even a complete absence of the effect was obtained for the cases with a ballistic spacer and $T_{_R}=0.5$--- when $\ar=10$ was considered. This is the reason why we chose the lower value of $\ar=5$ for the cases displayed in Fig.~\ref{fig:8}, in contrast to those presented in previous figures where $\ar=10$ seems to improve the TR effect. Our results are compatible with those obtained for weakly coupled lattices~\cite{He09} and those for homogeneous ones~\cite{He10} which indicate that NDTR mainly occurs in small-size systems, which is in line with the current trend of device miniaturization in the technological world~\cite{Cahill14}.

\section{Final remarks\label{sec:Disc}}

In this work we have performed the study of the rectification properties of a harmonic oscillator lattice coupled to substrates at both ends; first, in a way in which the unconnected central region (ballistic spacer) has a smaller length than that of its regions connected to the substrates at both ends, and afterwards when there is no central region at all. For all considered variations of both structural parameters ---mass-asymmetry magnitude and onsite potential amplitude--- and temperature bias magnitude the TR efficiency of the lattice without a ballistic spacer is higher than that of the corresponding instance with one. The results of the former case, for the employed system sizes considered, are better than those reported in Ref.~\cite{Guimaraes15}, which were obtained for a classical harmonic chain subject to an asymmetric quartic onsite potential, just as our model without a ballistic spacer, but with all oscillators subjected to an energy-conserving noise that randomly flips their velocity with a fixed rate. For example, the rectification for the case depicted in Fig.~\ref{fig:1}(a) with $\To=0.1$ and $\massl=8$ is $r=17.53$, or $0.89$ employing their definition $r=(J_+-|J_-|)/(J_++|J_-|)$, whereas the best results of the aforementioned work are, even in the small system-size limit, not better than $r\approx0.35$ with an average temperature value of $\To=1$. The better rectification value herein obtained can be ascribed to the added segmented-mass asymmetry, not considered in Ref.~\cite{Guimaraes15}, besides the one in the amplitudes of the onsite potential in both halves of the system. For the case when the ballistic channel is present this result entails the possibility of considering alternative mass distributions for the ballistic channel oscillators to explore possible rectification enhancements, such as the graded mass distribution considered in the harmonic 1D oscillator system of Ref.~\cite{Xiong19}, which has been recently applied to the amorphous interface between crystalline Si and Ge leads to manipulate the interfacial thermal conductance of the system~~\cite{Yang23}. In general, the reduction in TR efficiency for the system with a ballistic spacer seems to have it origin in the increased contribution of the high-frequency phonons in both the spacer and the leads, which entails an increase of the heat flux in the reverse-bias configuration that drastically reduces the rectification efficiency of the device. However, for an increase in the temperature bias the rectification values, for a low average temperature value, are almost insensitive to the increase on the size of the leads in the absence of the spacer, and only slightly dependent in its presence. Also, for the NDTR effect the best results are again obtained for low $\To$ values. For more complex systems it is also possible that asymmetric properties in the leads could also result in a reduction of TR if the ballistic channel has a shorter length than the leads; testing this hypothesis could be an interesting topic for future research.

For a possible experimental implementation there are a number of materials that have a high thermal conductivity, such as graphene~\cite{Xu14}, carbon nanotubes~\cite{Donadio07} or carbyne~\cite{Wang15} that could be considered as suitable candidates for a ballistic spacer. Then an asymmetry by means of defects, mass loading, or mechanical strains, among others, could be applied on its two ends in order to complete the implementation of the device. Nevertheless, it is worth mentioning that recently a gold-carbon nanotube system that is chemically bonded by molecular junctions has been theoretically studied~\cite{Dong19,Diao20} and shown to present significant TR. Now, given the significant mass gradients formed between gold and the molecular junctions or inside the elements of the latter, it can be hypothesized that this system could be considered as a practical realization of the herein considered 1D mass-graded system to achieve TR with currently available materials. To finish this work it is worth remarking that an enhancement of thermal conductivity can be obtained when the structure of the substrate is explicitly considered; for example, in coupled nanotubes~\cite{Guo11}, in graphene supported on Silicon dioxide~\cite{Ong11} as well as in a model of nonlinear 1D lattices coupled via van der Waals interactions~\cite{Sun13}. Therefore, there is the enticing possibility of manipulating the conductivity, and hence the rectification efficiency, of the herein considered system if the interaction with a more detailed substrate is considered. We will address this possibility in future work.

\ack
M.~R.~B. thanks Consejo Nacional de Humanidades Ciencias y Tecnolog\'\i as, M\'exico for financial support, Katheryn Serrano-Calabuche for her help in obtaining some of the results, and Maria~del~Carmen Nu\~nez-Santiago for useful comments and discussions. Both authors thank the anonymous referees that greately helped to improve our work.


\section*{References}

\providecommand{\newblock}{}

\end{document}